\documentclass[notitlepage,a4,article,nofootinbib,11pt]{article}

\usepackage{graphicx}

\usepackage{amsmath}%
\usepackage{amsfonts}%
\usepackage{amssymb}%
\usepackage{mathrsfs}
\usepackage{amsthm}

\usepackage{authblk}

\newcommand{\mc}{\mathcal}
\newcommand{\cp}{\times}

\newcommand{\cu}{\nabla\times} 
\newcommand{\JI}[1]{\bol{#1}\cdot\cu{\bol{#1}}}

\newcommand{\bol}{\boldsymbol}
\newcommand{\ti}{\textit}
\newcommand{\tb}{\textbf}

\newcommand{\abs}[1]{\left\lvert{#1}\right\rvert}
\newcommand{\w}{\wedge}
\newcommand{\lr}[1]{\left({#1}\right)}
\newcommand{\ea}[1]{\left\langle{#1}\right\rangle}

\newcommand{\mf}{\mathfrak}
\newcommand{\p}{\partial}

\newcommand{\ov}[1]{\mkern 1.5mu\overline{\mkern-1.5mu#1\mkern-1.5mu}\mkern 1.5mu}

\begin{document}
\title{Beltrami Operators}
\author[1]{N. Sato}
\affil[1]{Research Institute for Mathematical Sciences, \protect\\ Kyoto University, Kyoto 606-8502, Japan \protect\\ Email: sato@kurims.kyoto-u.ac.jp}
\date{\today}
\setcounter{Maxaffil}{0}
\renewcommand\Affilfont{\itshape\small}


\maketitle

\begin{abstract}
Beltrami fields occur as stationary solutions of the Euler equations of fluid flow and as force free magnetic fields in magnetohydrodynamics.
In this paper we discuss the role of Beltrami fields when considered as operators acting on a Hamiltonian function to generate particle dynamics.
Beltrami operators, which include Poisson operators as a special subclass, arise in the description of topologically constrained diffusion in non-Hamiltonian systems. Extending previous results \cite{Sato1}, we show that random motion generated by a Beltrami operator satisfies an H-theorem,
leading to a generalized Boltzmann distribution on the coordinate system where the Beltrami condition holds. 
When the Beltrami condition is violated, random fluctuations do not work anymore to homogenize the particle distribution in the coordinate
system where they are applied. The resulting distribution becomes heterogeneous. The heterogeneity depends on the `field charge' 
measuring the departure of the operator from a Beltrami field.    
Examples of both Beltrami and non-Beltrami operators in three real dimensions together with the corresponding equilibrium distribution functions are given.
\end{abstract}


\section{Introduction}
A 3-dimensional Beltrami field is a vector field aligned with its own curl. 
Beltrami fields arise as stationary solutions of hydrodynamic \cite{Moffatt,Enciso} and magnetohydrodynamic fluid equations \cite{Yos2002,Mah}.  
In these systems, a Beltrami field represents a physical field, 
such as fluid velocity or magnetic field. 
When the motion of a particle is considered,
these fields behave as antisymmetric operators that generate particle dynamics
by acting on the particle Hamiltonian. 

Antisymmetric operators, mathematically represented by bivectors, 
generalize Poisson operators \cite{Littlejohn1982,Morrison} of noncanonical Hamiltonian mechanics by allowing violation of the Jacobi identity \cite{Caligan}. 
This generalization becomes necessary when non-integrable constraints affect a dynamical system \cite{Bloch,Bates,Schaft}. 
An example pertaining to plasma physics is 
$\bol{E}\cp\bol{B}$ drift motion \cite{Cary}, where the magnetic field
plays the role of antisymmetric operator, and the electric field represents the
gradient of the Hamiltonian, in this case given by the electrostatic potential (in this paper we refer to particle energy as Hamiltonian function, even if the Jacobi identity is violated). 
Object of the present study are Beltrami operators, i.e. antisymmetric operators that satisfy the Beltrami condition.

If we consider an ensemble of particles endowed with an antisymmetric operator, particle interactions that
drive the system toward the equilibrium state can be represented by allowing a non-deterministic time-dependent part in the Hamiltonian function.
Then the equation of motion is stochastic in nature, and it can be translated into a Fokker-Planck equation for
a probability density (the distribution function of the ensemble, see refs. \cite{Gardiner,Risken,Sato1}).   
The resulting diffusion operator, which is written in terms of the antisymmetric operator,
is a second order non-elliptic partial differential operator. 
Violation of ellipticity (see refs. \cite{Gilbarg,Evans} for the definition of elliptic differential operator) 
occurs due to the null-space of the antisymmetric operator.
Such null-space, which reflects constraints affecting particle motion, makes the coefficient matrix
of the diffusion operator degenerate, thus breaking ellipticity.   

Integrability of constraints (in the sense of the Frobenius theorem \cite{Frankel}) is essential
in determining the geometrical properties of the antisymmetric operator:
according to the Lie-Darboux theorem of differential geometry \cite{Littlejohn1982,DeLeon_3, Arnold_4},
the null-space of any constant rank Poisson operator is locally and completely integrable in terms of Casimir invariants.
The level sets of the Casimir invariants define a symplectic submanifold where phase space coordinates
are available. Hence, the outcome of diffusion in a noncanonical Hamiltonian system
is a generalized Boltzmann distribution on the phase space metric of the symplectic submanifold
weighted by the Casimir invariants \cite{Yos2014,Sato2,Sato1}. 
 
When considering diffusion of an ensemble of particles in a given coordinate system and according to a prescribed antisymmetric operator, 
it has been shown in \cite{Sato1} that the Beltrami condition is the minimal requirement needed for the distribution function to homogenize (i.e. for the entropy to maximize) in that same reference frame. 
In the first part of this paper we partly generalize this result (which concerned pure diffusion processes) 
by allowing a deterministic part in the Hamiltonian function, as well as a friction term in the stochastic equation of motion,
and prove an H theorem. 
The result is a generalized Boltzmann distribution on the coordinate system where the Beltrami condition holds.
In the second part we restrict our attention to specific examples of pure diffusion processes in three real dimension,
and obtain the equilibrium distribution function for a set of both Beltrami and non-Beltrami operators. 
 
\section{Mathematical preliminaries}

We consider a smoothly bounded connected domain $\Omega\subset\mathbb{R}^{n}$ with boundary $\p\Omega$ and a 
coordinate system $\bol{x}=\lr{x^{1},...,x^{n}}$ with tangent basis $\lr{\p_{1},...,\p_{n}}$.
An antisymmetric operator is a bivector field $\mc{J}\in\bigwedge^2 T\Omega$ such that:
\begin{equation}
\mc{J}=\sum_{i<j}\mc{J}^{ij}\p_{i}\w \p_{j}=\frac{1}{2}\mc{J}^{ij}\p_{i}\w\p_{j},~~~~\mc{J}^{ij}=-\mc{J}^{ji}.
\end{equation}
In the following we assume that $\mc{J}^{ij}\in C^{\infty}(\ov{\Omega})$, $i,j=1,...,n$.
Given a Hamiltonian function $H\in C^{\infty}(\ov{\Omega})$, the equations of motion generated
by $\mc{J}$ can be written as a vector field $X\in T\Omega$:
\begin{equation}
X=\mc{J}\lr{dH}=\mc{J}^{ij}H_{j}\p_{i}.\label{X}
\end{equation} 
Due to antisymmetry we have:
\begin{equation}
\dot{H}=i_{X}dH=\mc{J}^{ij}H_{i}H_{j}=0.
\end{equation}
Here $i$ is the contraction operator. 
Since the Hamiltonian $H$ is preserved, $X$ is called a conservative vector field. Each $\mc{J}$ defines an antisymmetric bilinear form (bracket) $\left\{\cdot,\cdot\right\}$. Given a pair of smooth functions $f,g\in C^{\infty}(\ov{\Omega})$, we have:
\begin{equation}
\left\{f,g\right\}=\mc{J}\lr{df,dg}=f_{i}\mc{J}^{ij}g_{j}.
\end{equation}
The evolution of a function $f$ with respect to the flow \eqref{X} can be written as:
\begin{equation}
\dot{f}=\left\{f,H\right\}.
\end{equation}
$\mc{J}$ is called a Poisson operator whenever it satisfes the Jacobi identity:
\begin{equation}
h^{ijk}=\mc{J}^{im}\mc{J}^{jk}_{m}+\mc{J}^{jm}\mc{J}^{ki}_{m}+\mc{J}^{km}\mc{J}^{ij}_{m}=0,~~~~\forall~ i,j,k=1,...,n.
\end{equation}
The Jacobi identity is equivalent to demanding that the following trivector
vanishes identically:
\begin{equation}
\mf{G}=\frac{1}{2}\mc{J}^{im}\mc{J}_{m}^{jk}~\p_{i}\w \p_{j}\w \p_{k}=\sum_{i<j<k}h^{ijk}~\p_{i}\w \p_{j}\w \p_{k}.
\end{equation}
We shall refer to $\mf{G}$ as the Jacobiator, and to the tensor $h^{ijk}$
as the helicity density of $\mc{J}$.

We introduce an auxiliary volume form:
\begin{equation}
{\rm vol}_{g}=g~dx^{1}\w ... \w dx^{n},~~~~g\in C^{\infty}(\ov{\Omega}),~~~~g\neq 0.
\end{equation}
By contracting the bivector $\mc{J}$ with this volume form
we define the covorticity n-2 form:
\begin{equation}
\mc{J}^{n-2}=i_{\mc{J}}{\rm vol}_{g}=2\sum_{i<j}\lr{-1}^{i+j-1}g\mc{J}^{ij}dx^{n-2}_{ij}.
\end{equation}
In this notation $dx^{n-2}_{ij}=dx^{1}\w ... \w dx^{i-1}\w dx^{i+1}\w ... \w dx^{j-1}\w dx^{j+1}\w ... \w dx^{n}$.
The cocurrent n-1 form is defined to be:
\begin{equation}
\mc{O}^{n-1}=d\mc{J}^{n-2}=2\lr{-1}^{j}\frac{\p \lr{g\mc{J}^{ij}}}{\p x^{i}}~dx_{j}^{n-1}.
\end{equation}
The flow \eqref{X} admits an invariant measure ${\rm vol}_{g}$ for some appropriate metric $g$ and for any choice of $H$ provided that:
\begin{equation}
\mf{L}_{X}{\rm vol}_{g}=\frac{1}{g}\frac{\p \lr{g\mc{J}^{ij}}}{\p x^{i}}H_{j} {\rm vol}_{g}=0~~~~\forall H.
\end{equation} 
This is equivalent to demanding that $\mc{O}^{n-1}$ is identically zero.
Therefore, an antisymmetric operator satisfying $\mc{O}^{n-1}=0$ for some metric $g$ will be called measure preserving.

Due to the Lie-Darboux theorem, a constant rank Poisson operator
locally defines a symplectic submanifold. 
This submanifold is endowed with
the invariant measure provided by Liouville's theorem.
Hence, a constant rank Poisson operator is locally measure preserving.
Note that however not all measure preserving operators are locally Poisson.

Finally, consider the quantities:
\begin{subequations}
\begin{align}
b^{n-1}&=\mc{J}^{n-2}\w \ast d\mc{J}^{n-2},\label{FF}\\
\mf{B}&=\ast db^{n-1}.\label{FC}
\end{align}
\end{subequations}
We call \eqref{FF} the field force of $\mc{J}$ and \eqref{FC} its field charge.
An antisymmetric operator $\mc{J}$ will be called a \ti{Beltrami operator}
whenever $b^{n-1}=0$. If $\mf{B}=0$, the operator will be a \ti{weak Beltrami operator}. A Beltrami operator will be \ti{nontrivial} if $\mf{G}\neq \bol{0}$.
Note that in these definitions we do not require the Hodge star operator $\ast$ to be defined with respect to the same volume form ${\rm vol}_{g}$ used to define the operators. When $n=3$, $\lr{x^{1},x^{2},x^{3}}=\lr{x,y,z}$ is a Cartesian coordinate system, and $\ast$ is defined with respect to the Euclidean metric of $\mathbb{R}^3$,
the field force reduces to $\ast b^{1}=4\left[\bol{w}\cp\lr{\nabla\cp\bol{w}}\right]_{i}dx^{i}$, where $w_{x}=\mc{J}^{zy}$, $w_{y}=\mc{J}^{xz}$, $w_{z}=\mc{J}^{yx}$, and $\bol{w}=\lr{w_{x},w_{y},w_{z}}$. Thus, in $\mathbb{R}^3$ a Beltrami operator is nothing but a vector field satisfying the Beltrami condition $\bol{w}\cp\lr{\nabla\cp\bol{w}}=\bol{0}$. It is also worth oberving that, in $\mathbb{R}^3$, the Jacobi identity reduces to $h=\JI{w}=0$. Hence, a Beltrami operator is `dual' to a Poisson operator in the sense that while the former is aligned with its own curl, the latter is perpendicular to it.
Table \ref{Tab1} summarizes the geometrical quantities introduced in this section.

\begin{table}[h]
\footnotesize
\caption{List of geometrical quantities. AS stands for antisymmetric.}\label{Tab1}

\centering

\begin{tabular}{c c c c}
\\
\hline\hline
Name & Symbol & Definition & Expression\\
\hline\hline

Volume  & ${\rm vol}_{g}$ & & $g~ dx^{1}\w ... \w dx^{n}$\\

AS matrix  & $\mc{J}$ & & $\mc{J}^{ij}=-\mc{J}^{ji}$\\
 
AS operator  & $\mc{J}$ & &$ \frac{1}{2}\mc{J}^{ij}\p_{i}\w\p_{j}$\\

Helicity density  & $h^{ijk}$ & & $\mc{J}^{im}\mc{J}_{m}^{jk}+\mc{J}^{jm}\mc{J}_{m}^{ki}+\mc{J}^{km}\mc{J}_{m}^{ij}$\\

Jacobiator  & $\mf{G}$ & & $\frac{1}{2}\mc{J}^{im}\mc{J}_{m}^{jk}\p_{i}\w\p_{j}\w\p_{k}$\\

Covorticity  & $\mc{J}^{n-2}$ & $i_{\mc{J}}{\rm vol}$ & $2\sum_{i<j}\lr{-1}^{i+j-1}g\mc{J}^{ij}dx^{n-2}_{ij}$\\

Cocurrent & $\mc{O}^{n-1}$ & $d\mc{J}^{n-2}$ & $2\lr{-1}^{j}\frac{\p\lr{ g\mc{J}^{ij}}}{\p x^{i}}dx^{n-1}_{j}$\\

Field force & ${b}^{n-1}$ & $\mc{J}^{n-2}\w \ast d\mc{J}^{n-2}$ & $4\sum_{i<j}\lr{-1}^{i+j+k-1}~\cp~~~~~$\\
            & & & $~~~~~\cp~g\mc{J}^{ij}\frac{\p\lr{g\mc{J}^{lk}}}{\p x^{l}} dx^{n-2}_{ij}\w \ast dx_{k}^{n-1}$\\

Field charge & $\mf{B}$ & $\ast db^{n-1}$ & $$\\
\hline\hline
\end{tabular}
\end{table}

\section{An H-theorem}
In the following we restrict our attention to a Cartesian coordinate
system $\bol{x}=\lr{x^{1},...,x^{n}}$ with the standard Euclidean metric of $\mathbb{R}^{n}$. All qunatities and operations will be defined according to such metric.
We consider an ensemble of $N$ particles, at first not interacting, and each obeying
the equation of motion:
\begin{equation}
X_{0}=\mc{J}\lr{dH_{0}}.
\end{equation}
Here $H_{0}\in C^{\infty}(\ov{\Omega})$ is a time-independent Hamiltonian function and $\mc{J}$ a Beltrami operator.
In this case, the Beltrami condition $b^{n-1}=0$ reads:
\begin{equation}
b^{n-1}=4\lr{-1}^{i-1}\mc{J}^{ij}\frac{\p\mc{J}^{lj}}{\p x^{l}}dx_{i}^{n-1}=0.
\end{equation}
This implies:
\begin{equation}
\mc{J}^{ij}\frac{\p\mc{J}^{lj}}{\p x^{l}}=0,~~~~\forall~i=1,...,n.\label{BP}
\end{equation}
Next, we let particles interact with each other.
The energy of a single particle now has the form $H=H_{0}+H_{1}$, 
where $H_{0}=H_{0}\lr{\bol{x},t}$ is a deterministic component,
and $H_{1}=H_{1}\lr{\bol{x},t}$ a stochastic interaction term.
$H_{0}$ may include self-induced potentials, such as an electric potential
generated by electromagnetic interactions. 
In the following, we require $H_{0}$ to satisfy the condition:
\begin{equation}
\ea{\p_{t}H_{0}}=\int_{\Omega}f\p_{t}H_{0}\,dV=0.\label{H0t}
\end{equation}
Here $f\in C^{\infty}\lr{\Omega}$ is the distribution function of the ensemble
defined on the volume element $dV=dx^{1}\w ... \w dx^{n}$.  
Equation \eqref{H0t} states that the ensemble average of the rate of change in $H_{0}$
vanishes. Regarding the stochastic interaction term $H_{1}$, we demand that:
\begin{equation}
H_{1}=D^{1/2}~x^{i}~\Gamma_{i},~~~~i=1,...,n.\label{H1}
\end{equation}
Here $\bol{\Gamma}=\lr{\Gamma_{1},...,\Gamma_{n}}$ is an n-dimensional
Gaussian white noise random process, and $D>0$ a positive spatial constant (diffusion parameter). 
Equation \eqref{H1} implies that the stochastic force 
causing relaxation, $-\nabla H_{1}=-D^{1/2}~\bol{\Gamma}$, 
is homogeneous in the Cartesian coordinate system of $\mathbb{R}^{n}$. 
We further assume that such stochastic force is counterbalanced by
a friction (viscous damping) force, $-\gamma X_{0}$, where $\gamma$ is a positive spatial constant (friction or damping parameter).
In summary, the equation of motion of a particle in the ensemble now reads:
\begin{equation}
\begin{split}
X^{i}&=\mc{J}^{ij}\lr{H_{0j}+\gamma \mc{J}^{jk}H_{0k}+D^{1/2}\Gamma_{j}}-\kappa \mc{J}^{ij}_{j}\\
&=\lr{\mc{J}^{ij}-\gamma\mc{J}^{ik}\mc{J}^{jk}}H_{0j}+D^{1/2}\mc{J}^{ij}\Gamma_{j}-\kappa \mc{J}^{ij}_{j}.\label{XDiff}
\end{split}
\end{equation}
Here $\kappa >0$ is a positive spatial constant.
The term involving $\kappa$ represents a second damping term that is needed
because the antisymmetric operator $\mc{J}$ is not measure preserving (i.e. $\mc{J}^{ij}_{j}\neq 0$). 

According to the Stratonovich convention,
the stochastic differential equation \eqref{XDiff} translates into the
following Fokker-Planck equation \cite{Gardiner,Sato1}:
\begin{equation}
\begin{split}
\frac{\p f}{\p t}&=-\frac{\p}{\p x^{i}}\lr{fZ^{i}}\\
&=\frac{\p}{\p x^{i}}\left[-\lr{\mc{J}^{ij}-\gamma\mc{J}^{ik}\mc{J}^{jk}}H_{0j}f+\kappa \mc{J}^{ij}_{j}f+\frac{1}{2}D\mc{J}^{ik}\frac{\p}{\p x^{j}}\lr{\mc{J}^{jk}f}\right].\label{FPE0}
\end{split}
\end{equation}
Here we defined $Z^{i}$ to be the Fokker-Planck velocity of the system:
\begin{equation}
Z^{i}=\lr{\mc{J}^{ij}-\gamma\mc{J}^{ik}\mc{J}^{jk}}H_{0j}-\kappa\mc{J}^{ij}_{j}-\frac{1}{2f}D\mc{J}^{ik}\frac{\p}{\p x^{j}}\lr{\mc{J}^{jk}f}.
\end{equation}
If $\mc{J}$ satisfies the Beltrami condition \eqref{BP}, the Fokker-Planck equation
\eqref{FPE0} can be simplified to:
\begin{equation}
\frac{\p f}{\p t}=\frac{\p}{\p x^{i}}\left\{f\left[-\mc{J}^{ij}H_{0j}+\kappa\mc{J}^{ij}_{j}+\frac{1}{2}D\mc{J}^{ik}\mc{J}^{jk}\lr{\frac{\p\log{f}}{\p x^{j}}+\frac{2\gamma}{D}H_{0j}}\right]\right\}.\label{FPE}
\end{equation}
The goal of the remaining part of this section is to to show that the Fokker-Planck equation \eqref{FPE} maximizes the entropy functional:
\begin{equation}
S=-\int_{\Omega}f\log f\,dV,
\end{equation}
i.e. we wish to show that for $t\geq 0$:
\begin{equation}
\frac{dS}{dt}\geq 0.\label{dSdt}
\end{equation}
Observe that this statement is not trivial because
the system under consideration is not Hamiltonian. 
Therefore Liouville's theorem does not hold, 
and there is no phase space measure 
that can be used to naturally induce an entropy functional.

In order to obtain \eqref{dSdt} we need some boundary conditions that
ensure the closure of the system. We assume $Z\cdot N=\lr{X_{0}-\kappa\mc{J}^{ij}_{j}\p_{i}}\cdot N=0$ on $\p\Omega$, where $N$ is the outward normal to the boundary.
Then, using the antisymmetry of $\mc{J}$, the Beltrami property \eqref{BP}, the Fokker-Planck equation \eqref{FPE}, and the boundary conditions to eliminate surface integrals, 
the rate of change in $S$ is:
\begin{equation}
\begin{split}
\frac{dS}{dt}&=-\int_{\Omega}\p_{t}f\lr{1+\log{f}}\,dV\\
             &=-\int_{\Omega}fZ^{i}\frac{\p\log f}{\p x^{i}}\, dV\\
						 &=
\int_{\Omega}\left[\mc{J}^{ij}_{i}H_{0j}f+\frac{1}{2}D\mc{J}^{ik}\mc{J}^{jk}\lr{\frac{2\gamma}{D}H_{0j}+\frac{\p\log{f}}{\p x^{j}}}f_{i}\right]\, dV.\label{dSdt2}
\end{split}
\end{equation}
We define $\beta=2\gamma/D$. After some manipulations:
\begin{equation}
\begin{split}
\frac{dS}{dt}&=\lr{1-\beta\kappa}\int_{\Omega}{f\mc{J}_{i}^{ij}H_{0j}}\,dV+\beta\kappa\int_{\Omega}{f\mc{J}_{i}^{ij}H_{0j}}\,dV+\\
             &~~~+\frac{1}{2}D\int_{\Omega}{f\mc{J}^{ik}\mc{J}^{jk}\lr{\beta H_{0i}+\frac{\p\log f}{\p x^{i}}}\lr{\beta H_{0j}+\frac{\p\log f}{\p x^{j}}}}\,dV+\\
						 &~~~-\frac{1}{2}D\int_{\Omega}{f\mc{J}^{ik}\mc{J}^{jk}\lr{\beta H_{0j}+\frac{\p\log f}{\p x^{j}}}\beta H_{0i}}\,dV.\label{dSdt3}
\end{split}
\end{equation}
On the other hand, conservation of total energy $E=\int_{\Omega}{fH_{0}}\,dV$ implies that:
\begin{equation}
\frac{dE}{dt}=\int_{\Omega}{\p_{t}f H_{0}+f\p_{t}H_{0}}\,dV=\int_{\Omega}fZ^{i}H_{0i}\,dV=0.
\end{equation}
Here we used equation \eqref{H0t}.
Substituting the Fokker-Planck velocity $Z$ into the equation above,
we arrive at the condition:
\begin{equation}
\frac{1}{2}D\int_{\Omega}{f\mc{J}^{ik}\mc{J}^{jk}\lr{\beta H_{0j}+\frac{\p\log{f}}{\p x^{j}}}H_{0i}}\,dV=
\beta\kappa\int_{\Omega}{f\mc{J}^{ij}_{i}H_{0j}}\,dV.\label{Et2}
\end{equation}
Substituting equation \eqref{Et2} into
\eqref{dSdt3} gives:
\begin{equation}
\frac{dS}{dt}=\lr{1-\beta\kappa}\int_{\Omega}{f\mc{J}^{ij}_{i}H_{0j}}\,dV+
\frac{1}{2}D\int_{\Omega}{f\abs{\mc{J}\lr{d\log{f}+\beta dH_{0}}}^2}\,dV.
\end{equation}
Hence, upon setting $\kappa=\beta^{-1}$, we arrive at:
\begin{equation}
\frac{dS}{dt}=\frac{1}{2}D\int_{\Omega}{f\abs{\mc{J}\lr{d\log{f}+\beta dH_{0}}}^2}\,dV\geq 0.
\end{equation}
At $t\rightarrow\infty$, we must have $dS/dt=0$. It follows that, if $f>0$, the distribution function satisfies:
\begin{equation}
\lim_{t\rightarrow\infty}\mc{J}\lr{d\log{f}+\beta dH_{0}}=0.\label{Jdf}
\end{equation}
Equation \eqref{Jdf} is a generalized Boltzmann distribution.
Indeed, if the matrix $\mc{J}^{ij}$ is invertible, equation \eqref{Jdf}
reduces to the standard Boltzmann distribution $f^{\infty}\propto\exp\left\{-\beta H_{0}^{\infty}\right\}$, where $f^{\infty}=\lim_{t\rightarrow\infty}f$ and $\beta H_{0}^{\infty}=\lim_{t\rightarrow\infty}\beta H_{0}$.
When the matrix $\mc{J}^{ij}$ has a null-space that is at least partially  integrable by some invariants $C^{i}$, i.e. $\mc{J}\lr{dC^{i}}=\bol{0}$, one has $f^{\infty}\propto\exp\left\{-\beta H_{0}^{\infty}-\mu\lr{C^{i}}\right\}$, with $\mu$ a function of the invariants $C^{i}$ depending on the initial configuration of the system. Remember that if $\mc{J}$ is a constant rank Poisson operator, its null-space is always completely locally integrable in terms of Casimir invariants (see \cite{Sato1}).

\section{Examples in three real dimensions}
In $\mathbb{R}^{3}$ the action of $\mc{J}$ on a function $H$ can be
represented as a cross product:
\begin{equation}
\mc{J}\lr{dH}=\bol{w}\cp\nabla H.
\end{equation}
Here $\bol{w}\in C^{\infty}(\ov{\Omega})$ is the smooth vector field 
encountered in section 2.
When written in terms of $\bol{w}$, the Fokker-Planck equation \eqref{FPE} 
takes the form:
\begin{equation}
\frac{\p f}{\p t}
=\nabla\cdot\left\{\bol{w}\cp\left[\lr{-\nabla H_{0}+\gamma\,\nabla H_{0}\cp\bol{w}}f+\frac{1}{2}D\,\nabla\cp\lr{\bol{w} f}\right]-\kappa\, f\,\nabla\cp\bol{w}\right\}.\label{FPEw}
\end{equation}
Assume that $\bol{w}\neq\bol{0}$ in $\Omega$.
Then, if $\bol{w}$ is Beltrami field, $\nabla\cp\bol{w}=\hat{h}\,\bol{w}$ with $\hat{h}=\lr{\JI{w}}/\bol{w}^2$. Recalling that for a Beltrami operator $\kappa=\beta^{-1}=D/2\gamma$, equation \eqref{FPEw} can be written as:
\begin{equation}
\begin{split}
\frac{\p f}{\p t}
&=-\frac{\hat{h}}{\beta}\nabla\lr{\beta H_{0}+\log{f}}\cdot\bol{w}\, f+
\nabla H_{0}\cdot\bol{w}\cp\nabla f
+\\
&~~~+\frac{1}{2}D\,\nabla\cdot \left\{\left[\bol{w}\cp\nabla\lr{\beta\, H_{0}+\log{f}}\cp\bol{w}\right] f\right\}.\label{FPEw2}
\end{split}
\end{equation}

In the following we restrict our attention to stationary solutions to
the purely diffusive form of \eqref{FPEw}, which is obtained by setting $H_{0}=\kappa=0$. In the limit $t\rightarrow\infty$ we have $\p_{t}f=0$. Therefore:
\begin{equation}
0=\nabla\cdot\left\{\bol{w}\cp\left[\nabla\cp\lr{\bol{w} f}\right]\right\}.\label{Diff}
\end{equation}
Here we have written $f$ in place of $f^{\infty}$ to simplify the notation.
This convention will be used in the rest of the paper.

The field force associated to $\bol{w}$ can be expressed in vector form
as $\bol{b}=\bol{w}\cp\lr{\nabla\cp\bol{w}}$. Then the field charge reads $\mf{B}=\nabla\cdot\bol{b}$. In terms of $\bol{b}$ and $\mf{B}$ equation \eqref{Diff} becomes:
\begin{equation}
0=\mf{B}f+\bol{b}\cdot\nabla f+\nabla\cdot\left[\bol{w}\cp\lr{\nabla f\cp\bol{w}}\right].\label{Diff2}
\end{equation}
Let us now study some explicit examples.

\subsection{Example 1: Beltrami operators in $\mathbb{R}^3$}
From equation \eqref{Diff2} it is clear that
any weak Beltrami operator ($\mf{B}=0$) admits the stationary solution $\nabla f=\bol{0}$. This solution is also the outcome of the dynamical diffusion process for any nontrivial Beltrami operator, as follows from the proof of the H theorem. Indeed if the operator is nontrivial, $\mf{G}\neq\bol{0}$,
implying $h=\JI{w}\neq 0$. This means that the Frobenius integrability condition
for the vector field $\bol{w}$ is violated, and there is no function $C$ such that $\bol{w}\cp\nabla C=\bol{0}$. Hence the only solution to \eqref{Jdf} is $\nabla f=\bol{0}$. Conversely, if $\mf{B}\neq 0$ such solution is not admissible. 

The classical Beltrami field in $\mathbb{R}^3$ has the form:
\begin{equation}
\bol{w}=\sin{z}\,\nabla x+\cos{z}\,\nabla y.\label{B1}
\end{equation}
This vector field satisfies $\nabla\cp\bol{w}=\bol{w}$, $\nabla\cdot\bol{w}=0$, and $\bol{w}^2=1$.
Let $\sigma=\sigma\lr{z}$ be a smooth function of the variable $z$.
We can slightly generalize equation \eqref{B1} as:
\begin{equation}
\bol{w}=\sin{\sigma}\,\nabla x+\cos{\sigma}\,\nabla y,\label{B2}
\end{equation}
which satisfies $\nabla\cp\bol{w}=\sigma_{z}\,\bol{w}$, $\nabla\cdot\bol{w}=0$, and $\bol{w}^2=1$. 
Notice that the vector fields \eqref{B1} and \eqref{B2} are solenoidal. Hence they represent stationary solutions of the ideal Euler equations of fluid dynamics at constant density and with pressure $P=-\bol{w}^2/2$.

More generally, suppose that $\lr{\ell,\psi,\theta}$ is an orthogonal coordinate system such that $\abs{\nabla\ell}=\abs{\nabla\psi}$.
Consider the vector field:
\begin{equation}
\bol{w}=\cos{u}\,\nabla\psi+\sin{u}\,\nabla\ell,\label{B3}
\end{equation}
where $u=u\lr{\theta}$ is a smooth function of the variable $\theta$.
We have:
\begin{equation}
\nabla\cp\bol{w}=u_{\theta}\,\nabla\ell\cdot\nabla\psi\cp\nabla\theta\lr{\sin{u}\,\frac{\nabla\ell}{\abs{\nabla\ell}^2}+\cos{u}\,\frac{\nabla\psi}{\abs{\nabla\psi}^2}}=u_{\theta}\,\abs{\nabla\theta}\,\bol{w}.
\end{equation}
If $u_{\theta}\neq 0$, this equation also implies that:
\begin{equation}
\bol{w}=u_{\theta}^{-1}\abs{\nabla\theta}^{-1}\nabla\theta\cp\nabla\lr{\ell\,\cos{\theta}-\psi\,\sin{\theta}}.
\end{equation}
Hence, if we interpret $\bol{w}$ as a flow, it has two integral invariants $\theta$ and $\ell\,\cos{\theta}-\psi\,\sin{\theta}$.

For example, take $\lr{\ell,\psi,\theta}=\lr{\sqrt{\rho+z},\sqrt{\rho-z},\arctan\lr{y/x}}$ to be a parabolic coordinate system with $\rho^2=x^2+y^2+z^2$.
This coordinate system is orthogonal. Furthermore $\abs{\nabla\ell}^2=\abs{\nabla\psi}^2=1/2\rho$ and $\abs{\nabla\theta}=1/\sqrt{x^2+y^2}$. Hence, the vector field:
\begin{equation}
\bol{w}=\cos\left[{\arctan\lr{\frac{y}{x}}}\right]\,\nabla\sqrt{\rho-z}+\sin\left[\arctan\lr{\frac{y}{x}}\right]\,\nabla\sqrt{\rho+z},\label{B4}
\end{equation}
is a Beltrami field with proportionality factor $\hat{h}=1/\sqrt{x^2+y^2}$, i.e. $\nabla\cp\bol{w}=\hat{h}\,\bol{w}$.
In a similar way we can construct arbitrarily complex Beltrami fields
by finding appropriate orthogonal coordinate systems. 

More precisely, suppose that we want a Beltrami field
with a given proportionality coefficient $\hat{h}$ such that $\nabla\cp\bol{w}=\hat{h}\,\bol{w}$. This can be accomplished by first solving the Eikonal equation $\abs{\nabla\theta}=\lvert{\hat{h}\rvert}$ and then by trying to determine the orthogonal coordinate system $\lr{\ell,\psi,\theta}$ with respect to the obtained $\theta$ (it is essential that $\abs{\nabla\ell}=\abs{\nabla\psi}$). 
 Notice that however, while the Eikonal equation for the variable $\theta$ can be solved within the framework of the method of characteristics for first order partial differential equations, the existence of the coordinate system $\lr{\ell,\psi,\theta}$ is not guaranteed in general.

As an example, take $\hat{h}=\exp\lr{x+y}$. Applying the procedure described above,
one can construct the Beltrami field:
\begin{equation}
\bol{w}=\frac{1}{\sqrt{2}}\cos\left[\frac{\exp\lr{x+y}}{\sqrt{2}}\right]~\nabla \lr{x-y}+\sin\left[\frac{\exp\lr{{x+y}}}{\sqrt{2}}\right]~\nabla z.
\end{equation}
Observe that this vector field is solenoidal. If one wants a Beltrami field with opposite proportionality factor $-\hat{h}$, it is sufficient to consider the dual vector field $\bol{w}^{\ast}=\sin{\theta}~\nabla\psi+\cos{\theta}~\nabla\ell$. 

Next, let us find a weak Beltrami operator, i.e. a vector field such that $\bol{b}\neq\bol{0}$ but $\mf{B}=0$. Consider the vector field:
\begin{equation}
\bol{w}=\sqrt{y^2-2 z^2}\,\nabla x+z\,\nabla y.\label{B5}
\end{equation}
This vector field satisfies $\bol{b}=\frac{1}{2}\nabla\lr{y^2-z^2}-\frac{zy}{\sqrt{y^2-2z^2}}\nabla x$ and $\mf{B}=0$ as desired.


\subsection{Example 2: non-Beltrami operators in $\mathbb{R}^3$}
Suppose that we can find a smooth function $g\in C^{\infty}(\ov{\Omega})$, $g\neq 0$, such that the Beltrami condition is satisfied by the vector field $g\,\bol{w}$, i.e.: 
\begin{equation}
g\bol{w}\cp\left[\nabla\cp\lr{g\bol{w}}\right]=\bol{0}.\label{NB1}
\end{equation}
Then $g$ is a stationary solution to \eqref{Diff}.
Observe that now $\bol{b}=-\bol{w}\cp\lr{\nabla\log{g}\cp\bol{w}}$.
Hence $\bol{w}$ is not a Beltrami operator in $\mathbb{R}^3$.
Condition \eqref{NB1} can be obtained by using
the auxiliary volume element ${\rm vol}_{g}=g\,dx\w dy\w dz$
while keeping the Hodge star on $\mathbb{R}^3$ 
in the definition of $b^{n-1}$, and by setting $b^{n-1}=0$.
It follows that the proper entropy measure for the diffusion process is $S=-\int_{\Omega}f\log\lr{\frac{f}{g}}\,dV$. Then, by the H theorem, $\bol{w}\cp\nabla \lr{f/g}\rightarrow \bol{0}$ in the limit $t\rightarrow\infty$. If $\mf{G}\neq\bol{0}$ this implies $f\rightarrow g$.

There is a class of vector fields that always satisfy equation $\eqref{NB1}$ 
for some appropriate choice of the function $g$.
Let $\lr{\ell,\psi,\theta}$ be the orthogonal coordinate system introduced in the previous example.
Let $u=u\lr{\theta}$ be a smooth function of the variable $\theta$.
Consider the vector fields:
\begin{equation}
\bol{w}=\nabla\psi+u\,\nabla\ell.\label{NB2}
\end{equation}
It can be verified that, by setting $g=1/\sqrt{1+u^2}$, condition \eqref{NB1}
is satisfied. Indeed, if we define a new variable $\sigma=\arctan{u}$ 
we obtain $g\,\bol{w}=\cos{\sigma}\,\nabla\psi+\sin{\sigma}\,\nabla\ell$, which has the same form of the Beltrami field \eqref{B3}.
 
A simple example of \eqref{NB2} is:
\begin{equation}
\bol{w}=\nabla x+y\,\nabla z.
\end{equation}
This vector field satisfies $\bol{b}=y\,\nabla y$ and $\mf{B}=1$.
The outcome of the corresponding diffusion process is the equilibrium distribution function $f=1/\sqrt{1+y^2}$.


\section{Concluding remarks}

Beltrami operators are bivectors that satisfy the Beltrami condition.
This class of operators allows a formulation of statistical mechanics
even in the absence of canonical phase space, i.e. even if the underlying dynamics is not Hamiltonian.
In this paper we studied the properties of Beltrami operators
in the context of statistical mechanics of topologically constrained
mechanical systems. We proved an H theorem for Beltrami operators
in $\mathbb{R}^n$, and obtained the equilibrium distribution function,
a generalized Boltzmann distribution. 
Examples of both Beltrami and non-Beltrami operators were given together with the resulting equilibrium distribution function for the case of pure diffusion processes in $\mathbb{R}^3$.

\section*{Acknowledgments} The research of N. S. was supported by JSPS KAKENHI Grant No. 18J01729. The author would like to acknowledge useful discussion with Professor Z. Yoshida and Professor M. Yamada.


\end{document}